\documentclass[3pt,onecolumn]{article}
\usepackage{graphicx}

\newcommand{\ben}{\begin{enumerate}}
\newcommand{\een}{\end{enumerate}}

\newcommand{\be}{\begin{equation}}
\newcommand{\ee}{\end{equation}}
\newcommand{\bea}{\begin{eqnarray}}
\newcommand{\eea}{\end{eqnarray}}

\begin{document}
\begin{flushright}
\end{flushright}
\vspace{0.1cm}
\thispagestyle{empty}

\begin{center}
{\Large\bf 
The ancient art of laying rope}\\[13mm]
{\rm Jakob Bohr$^*$ and Kasper Olsen}\\[2.5mm]
{\it Department of Physics,\\ Technical University of Denmark}\\
{\it Building 307 Fysikvej, DK-2800 Lyngby, Denmark}

{\tt jakob.bohr@fysik.dtu.dk, kasper.olsen@fysik.dtu.dk}\\[18mm]

{\bf Abstract}
\end{center}
\noindent {\bf We describe a geometrical property of helical structures and show how it accounts for the early art of ropemaking. Helices have a maximum number of rotations that can be added to them -- and it is shown that this is a geometrical feature, not a material property. This geometrical insight explains why nearly identically appearing ropes can be made from very different materials and it is also the reason behind the unyielding nature of ropes. The maximally rotated strands behave as zero-twist structures. Under strain they neither rotate one or the other way. The necessity for the rope to be stretched while being laid, known from Egyptian tomb scenes, follows straightforwardly, as does the function of the top, an old tool for laying ropes. 
}\\[5mm]


\addtocounter{section}{1}

The crafting of rope, string and cordage has been an essential skill through the times going back to early prehistoric life. The image of rope is easily discernible and could perhaps be said to be iconic. It has also been important for early symbolic meaning and for creed. Examples are  the spinning seidh \cite{heide2006}, the shimenawa prayer rope \cite{grim1982}, the shen ring, and the cartouche in hieroglyphs \cite{james1982}. Scenes from Egyptian tombs display advanced ropemaking \cite{teeter1987,charlton1996}. Figure 1 shows a scene from the tomb of Akhethotep and Ptahhotep, and it appears to be depicted that the ropes are held under tensile stress while being laid. The round tool hanging from the rope is perhaps a stone helping the ropemakers to gauge that sufficient tensile stress is present; to maintain a nearly straight rope requires the presence of adequate tensile strength depending on the weight of the stone.
In a scene from the tomb of Rekh-mi-Re  a special belt is depicted which help the ropemakers to apply tensile stress by the use of their body weight \cite{charlton1996}. Large quantities of ancient Egyptian rope has been found in a cave at the Red Sea coast \cite{veldmeijer2008}, also found at the site are two limestones with holes now described as anchors \cite{zazzaro2007,zazzaro2010}. We believe that another possible use of these stones might have been as weights used during the rope production akin to the tomb scene depicted in Figure 1. 

\begin{figure}[h]\centering
\includegraphics[width=10.428cm]{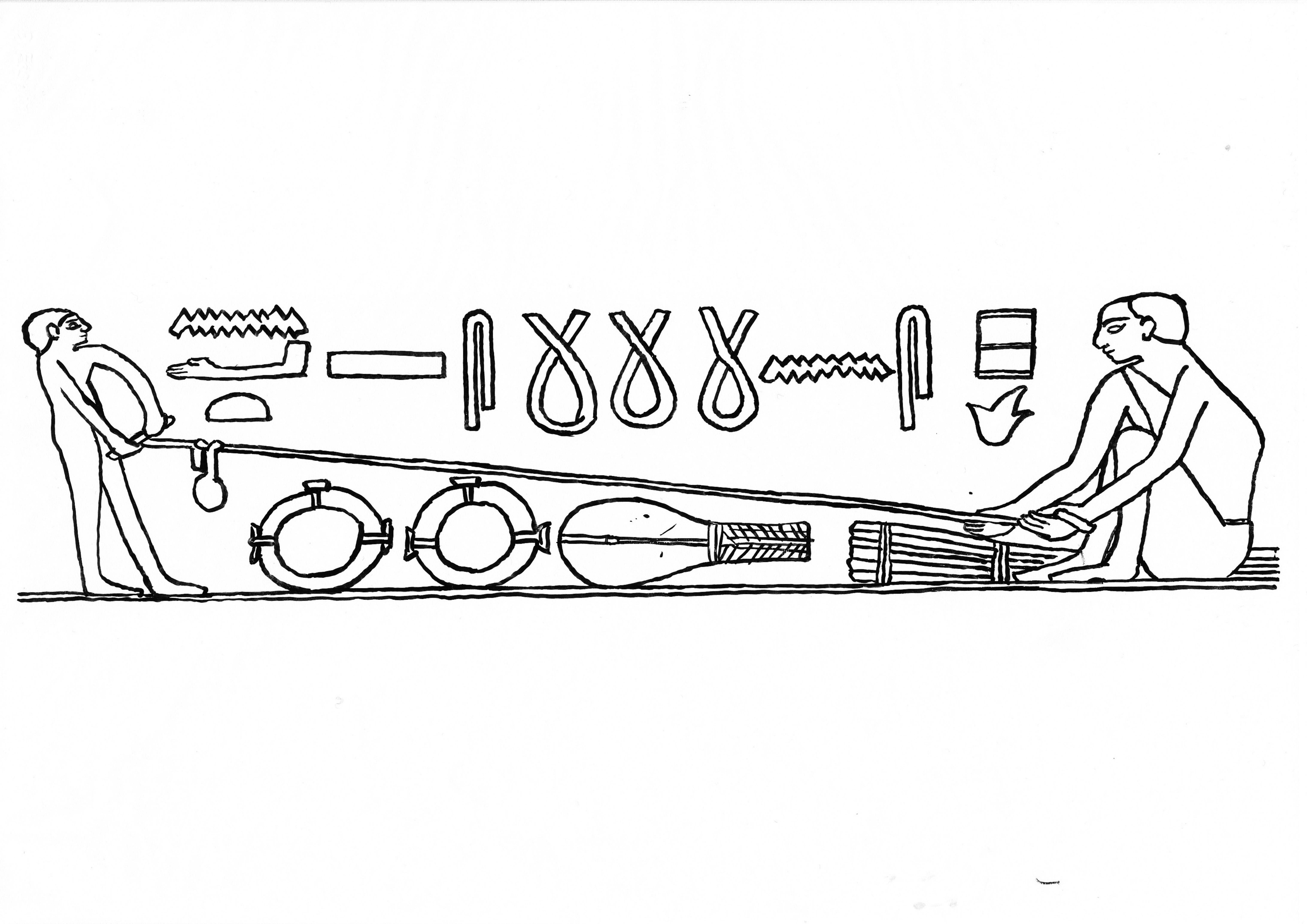}
\caption{\it Ropemaking in ancient egypt. Tomb of Akhethotep and Ptahhotep, about 2300 BC. The round tool hanging from the rope close to the person to the left is perhaps a stone helping the ropemakers to gauge that sufficient tensile stress is present.}
\end{figure}

Classical ropes appear with an easy discernible geometrical structure, even though they have been fabricated in different human cultures from a large variety of fibrous materials with diverse physical properties. Relatively recently, Zhang et al. has demonstrated that yarn formation can be performed with the use of nano-sized strands as well \cite{zhang2004}. Why does the resulting geometry of rope appear so similar, as if it depends little on the material used, and why are ropes inextensible? Here we show, that these properties of rope are due to a universal behavior of helical structures which depends on geometry. It stems from the observation, derived below, that there is a maximum number of rotations that can be added to a helical structure (or more precisely an $N$-helix, where $N \ge 2$). One consequence is that a tightly laid rope, where each of the strands are rotated to their maximum in one direction while being helically laid with the maximum number of rotations in the opposite direction will be interlocked, unable to unwind, and hence a functional rope. 

Mathematical aspects of the helical geometry of yarns are described by Treloar \cite{treloar1956} and by Fraser et al. \cite{fraser1998}, and a comprehensive review for wire rope is given by Costello \cite{costello1997}. The counter-twisting of the strands and the rope, respectively, has been discussed from a mathematical perspective \cite{fraser1998}, and ropemaking in a historical context 
\cite{mcgee1897}. From the point of view of topology it is important to consider the amount of writhing, see Thompson and Campneys \cite{thompsom1996} and Stump et al. \cite{stump1998}. Thompson et al. have suggested that double helices will kinematically lock-up at $45^\circ$ \cite{thompson2002}, Gonzales and Maddocks have introduced the {\it global curvature} and investigated the significance hereof when understanding helical structure formation \cite{gonzalez1999}. Przyby\l ~and Piera\'{n}ski have derived the conditions for self-contacts for single helices \cite{przybyl2001}, and Neukirch and van der Heijden the condition for inter-strand contacts in an $N$-ply \cite{neukirch2002}. Olsen and Bohr have determined the close-packed helical structures from a calculation of the volume fraction for a helix as a function of the pitch angle \cite{olsen2009}.

It is surprisingly simple to see that there is a geometrical limit to the number of rotations on a helix, 
in the following we show that helical structures can be maximally rotated.
A helical curve is uniquely described by two independent variables; in differential geometry it is common to use curvature and torsion. For a description applicable to ropes formed from tubular strands, we will use $(n_r, L_r)$ as parameters, where $n_r$ is the number of turns the helical strands makes on the imaginary cylinder of the rope, and $L_r$ is the length of the rope.  Not all combinations of $(n_r , L_r)$ are allowed as some are forbidden because of tubular interactions; the helical tubes are assumed to have hard walls and are therefore not allowed to intercept each other. The allowed values of $(n_r , L_r)$ are gray shaded in Figure~2. The solid line is the boundary case where the helical tubes are in contact with each other, i.e. where the distance between two neighboring strands in the rope is equal to the diameter, $D$, of one strand. Hence the solid line in Figure~2 corresponds to all the packed structures. If a rope is laid under tensile strength one obtains the structures that correspond to the upper part of the solid curve. The lower part corresponds to
strands which are laid out flat on top of each other and then twisted around an imaginary cylinder which radius becomes smaller and smaller as one progresses along the curve.

\begin{figure}[h]\centering
\includegraphics[width=13.428cm]{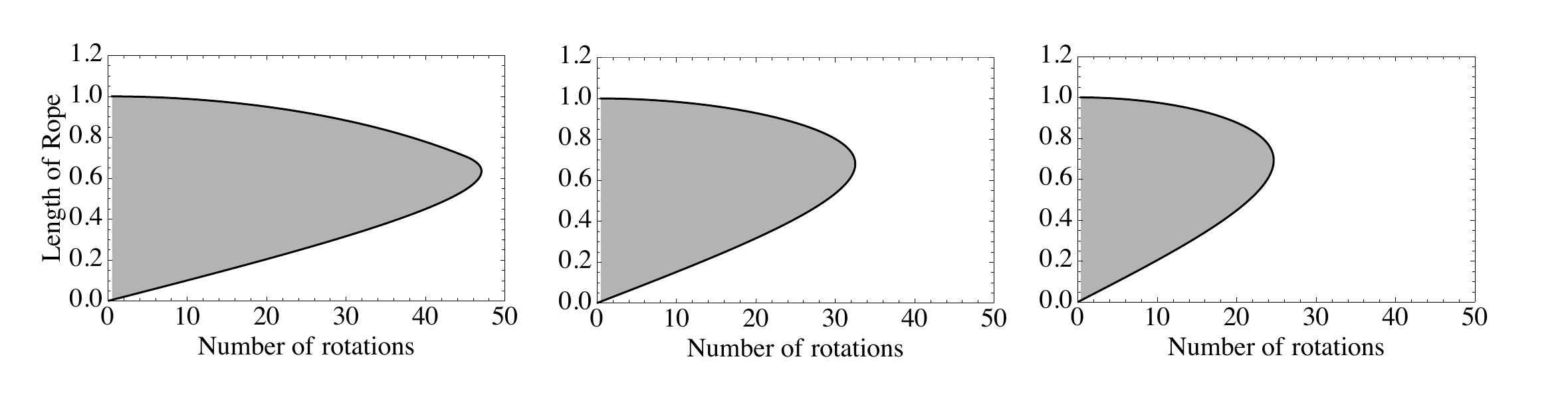}
\caption{\it The length of the formed rope as a function of the number of turns. The curve shown is for a two-, three- and four-stranded rope from left to right, respectively,. The ropes are formed from strands each being 1m long and having a strand diameter of 5~mm. The shape of these curves is universal for ropes with a given number of strands, while the specific number of rotations depend on the diameter and the length of the rope.
At the zero-twist point of the triple-stranded rope, the length of the formed rope is always 68\% of the length of the individual strands. For a two- or four-stranded rope the numbers are 63\% and 69\%, respectively.}
\end{figure}

Now, let us explain in detail how Figure~2 is obtained. For the calculation of the solid line it is useful to notice that the number of turns, $n_r$, can be expressed as

\begin{equation}
\label{eq1}
n_r=(L_s /2 \pi a) \cos v_\bot\, .
\end{equation}

\noindent Here $L_s$ is the length of the strands, $v_\bot$ is the pitch angle of the strands defined relative to the equatorial plane, and $a$ is the radius of an imaginary cylinder surface hosting the helical center line of the strands. The length of the fabricated rope, $L_r$, is

\begin{equation}
\label{eq2}
L_r=L_s \sin v_\bot\, .
\end{equation}

\noindent Equation (\ref{eq2}) contains the radius, $a$, of the imaginary cylinder hosting the helical lines. The radius $a$, which is not a constant as it depends on $v_\bot$, can be determined from the requirement that the strands are in contact with each other. Recently, this requirement has been solved for tubular helices \cite{przybyl2001,neukirch2002,olsen2009}. The specific solutions for the equations here are obtained through the transcendental equation:

\begin{equation}
\sin t +(\frac{L_r}{2 \pi n a})^2 (t+\frac{2\pi}{N}) = 0\, ,
\end{equation}

\noindent where $t$ is the parametric distance, $t=t_{i+1} -t_i$, between nearest points on neighboring helical curves
describing the strands (see e.g. \cite{olsen2009}).
%
The radius $a$ is found by the requirement that the nearest distance equals the strand diameter, $D$. I.e. through the condition that
\begin{equation}
D^2=a^2(\cos t -1)^2 + a^2\sin^2 t + (\frac{L_r}{2\pi n})^2(t+\frac{2\pi}{N})^2\, .
\end{equation}

\noindent In Figure 2 (where $N=2,3,4$) a striking result is immediately visible, namely that there is a maximum number of turns that one can have on a strand of a certain length! 

The peculiar point on the curve in Figure 2 where the maximum possible number of rotations is obtained is at the turning point of $L_r(n_r)$. At this point it is impossible for the strand to be further twisted and we henceforth denote this point as {\it zero-twist}. The zero-twist structure of an $N$-ply has the property that stretching  will neither make the individual strands rotate in one, nor in the other direction. This follows from the tangent of the curve being vertical and hence $dn_r/dL_r=0$. The total twist, $\Theta$, is the angular rotation of the strand around the imaginary cylinder, i.e. $\Theta/2\pi= n_r$, and therefore we also have, $d\Theta/dL_r=0$ for this particular number of rotations.

\begin{table*}[h]
\begin{tabular}{lllllll}
\hline\noalign{\smallskip}
No. of strands & $1$ & $2$ & $3 $ & $4$ & $\infty$\\
\noalign{\smallskip}\hline\noalign{\smallskip}
$v_{ZT}$ &$ - $ & $39.4^\circ$ & $42.8^\circ$& $43.8^\circ$ & $45^\circ$\\
\noalign{\smallskip}\hline
\end{tabular}
\caption{{\it Pitch angle, $v_{ZT}$, for the zero-twist structures given as a function of the number of strands. To these zero-twist helical structures one cannot add additional rotations. 
A rope with three strands where the individual strands each have two substrands is a typical classical rope.}}
\end{table*}

Generally, we will denote any structure that has a vanishing strain-twist coupling as being a zero-twist structure. For $N \ge 2$ the geometrical restrictions are given in Table 1 in terms of the pitch angles for zero-twist structures. The case $N=1$ is special as there is a divergence of the geometrical solution into two branches and therefore no universal geometrical point. If non-vanishing elastic constants are considered the divergence disappears and hence there is a material-dependent maximum number of turns that can be applied in the $N=1$ case. Figure 3 display the $N$-helix ($N=2,3,4$) with a pitch angle of the zero-twist structures as a mathematical idealization. 

\begin{figure}[h]\centering
\includegraphics[width=10.428cm]{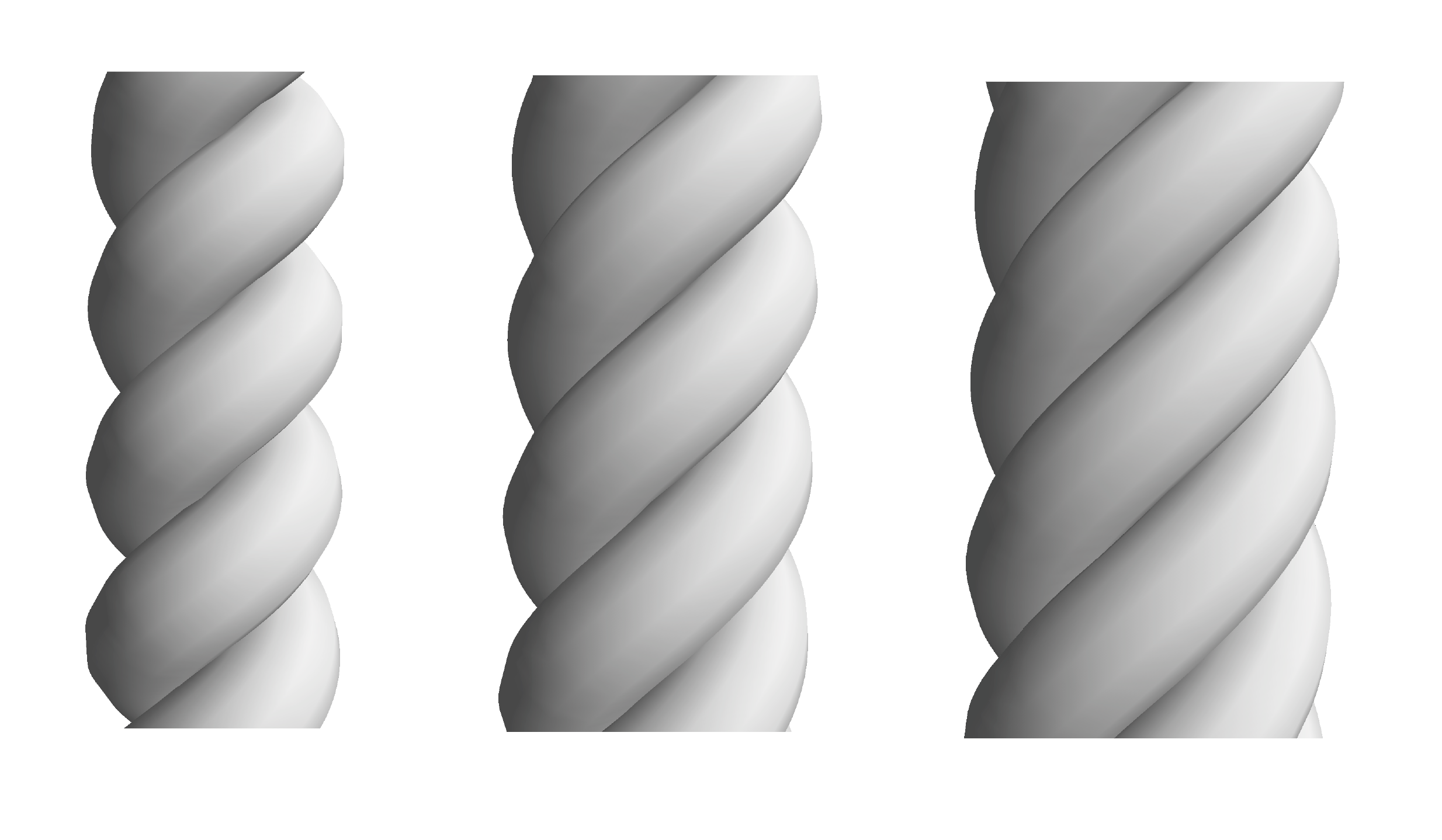}
\caption{\it An ideal representation of a two-, three- and four-stranded rope laid with a pitch angle corresponding to their zero-twist structures ($39.4^\circ$, $42.8^\circ$ and $43.8^\circ$ respectively relative to the equatorial plane). With these pitch angles, the strands will neither rotate in one or the other direction under vertical strain.}
\end{figure}

A perfectly tightly laid rope is a configuration where the strands are twisted to their zero-twist configuration in one direction while the rope is laid at its zero-twist in the opposite direction. Such a rope would be infinitely rigid when taken as an ideal tubular idealization with hard walls, i.e. it would have no flexibility. In practice even a tightly laid rope will have some flexibility.
Flexibility is introduced by laying the rope close to, but not exactly at the zero-twist configuration. When the rope is laid under tensile strength it will self-lock at the equilibrium
\begin{equation}
\label{equilibrium}
\frac{dL_r}{dn_r}=\frac{\partial L_r}{\partial n_r} +\frac{\partial L_r}{\partial L_s} \frac{\partial L_s}{\partial n_s} \frac{\partial n_s}{\partial n_r}=0\, ;
\end{equation}

\noindent  where $n_r$ is the number of turns the helical strands makes on the imaginary cylinder of the rope, and $n_s$ quantify the counter-rotation from the pre-twisting of the strands. For the tightly laid rope $n_r+n_s$ is equal to the sum of the maximum number of rotations (with sign) of the rope and of the strands, respectively. In general, $n_r+n_s = constant$, where the size of the constant determines how hard the rope is laid. A triple stranded rope where  $n_r$ is 94 \% of its maximum will have a pitch angle of 52$^\circ$. Notice, that it is  a geometrical property of the rope, and not a material property, that is responsible for the nature of the equilibrium given by Eq. (\ref{equilibrium}). This geometrical equilibrium of rope is what accounts for their unyielding nature (their ability to maintain a constant length).

A consequence of the presented analysis is that when laying the rope it is necessary to add the strands in a way that will allow pitch angles down to the one of the zero-twist configuration. According to Figure~2 the tensile stress in the rope will secure that the newly added incremental piece of rope be on the upper part of the curve on Figure~2, and if properly twisted it will then adjust itself towards the zero-twist structure. This is the purpose of the so-called top (a cone with grooves in) used at traditional ropewalks, and of similar tools. The top brings the strands together from a radius which is purposely too large. 

Simple steel-wires which are often laid with a somewhat higher pitch angle cannot be used in cranes with a single fall of the rope. The reason for this is that the load is not rotationally fixed when hanging and as a consequence the steel-wire would unwind. This has lead to the manufacturing of counter-rotated multi-layered steel wires which are denoted rotationally restricted \cite{pellow1982}. It is interesting to note that the classical ropes discussed above are relatively well performing rotationally restricted ropes, this being a consequence of the equilibrium described by Equation (\ref{equilibrium}). The need for this type of engineering therefore did not arise before the advent of the steel-wire; steel-wires are often modeled using mechanical concepts \cite{cardou1997}.

In summary, we have discussed the significance of the zero-twist geometry for ropemaking, i.e. the maximally rotated structures. 
In the hard-wall model used the physical forces that one strand is causing on another strand are implicitly implied. I.e. repulsions are assumed to be infinitely large when the hard-wall criterion is violated. This idealization of the forces is what allows for one to take a geometrical approach to understanding the helical structure of rope. At last, one can speculate why the described geometrical nature of the art of ropemaking has been overlooked. One explanation could be that by laying the rope through the usual procedures, it will -- for the reasons described above -- automatically be a functional rope. And, therefore the intrinsic geometry behind the art of laying rope is not something you have to know or be aware of, just the instructions which have been passed down through generations.

\subsection*{Acknowledgements}
We would like to thank P. Piera\'{n}ski for constructive comments on an early version of the manuscript. We would also like to thank M. J{\o}rgensen and T. Bagh at Ny Carlsberg Glyptotek, Copenhagen, for assistance with Fig. 1, and A. J. Veldmeijer and C. Zazzaro for helpful correspondence.


\noindent $^*$ Present address: Fuel Cell and Solid State Chemistry Division, Ris\o~National Laboratory for Sustainable Energy, Technical University of Denmark, 
DK-4000 Roskilde. E-mail: jabo@risoe.dtu.dk.

\end{document}